\documentclass[12pt,a4paper]{article}
\usepackage{hyperref}
\usepackage{amsmath}
\usepackage{amsfonts}
\usepackage{amssymb}
\usepackage{graphicx}
\usepackage{float}
\usepackage{mathtools,amscd}

\newcommand{\Tau}{\mathcal{T}}

\title{RR charge calculation for D brane in B field}
\author{Tetiana Obikhod, Ievgenii Petrenko}
\date{%
    \it{Kyiv Institute for Nuclear Research NAS of Ukraine}\\%
    \today
}

\begin{document}

\maketitle 

\section{Abstract}
\par
The paper is connected with searches for the Ramond-Ramond charge of D branes in the presence of B field. The consideration of B field inclusion is an important physical and mathematical unsolved problem, which is connected with K group calculations of twisted bundles. Considered two cases of vector bundles, Azumaya and Rosenberg algebras and analyzed their K group realization.  
\section{Introduction}
\par
One of the most interesting question of Ramond-Ramond(RR)-charge classification of D-brane is the description of corresponding vector bundles characterized by Dixmier—Douady invariant by twisted K-theory group \cite{Bouwknegt_2000}. As is known \cite{PhysRevLett.75.4724}, RR fields on D-branes are sources for fields of type II string theory. As quantum RR fields are classified by twisted K-theory, we'll present the mathematical consideration of RR-charge in terms of $C^*$-algebra on Hilbert space and corresponding topological invariant, element of twisted K-group.
\par
Historically, RR-charge appeared in the type II closed superstring theory with gauge fields from RR sectors of string Hilbert space, \cite{doi:10.1002/zamm.19880680630}. Inclusion of boundary conditions on open string endpoints leads to hyperplane, the D-brane, with p spatial and one timelike dimension. The quantum charge calculation of D-brane, includes the exchange of open and closed string between the two D-branes. From minimum quantum, $n=1$, it was argued that D branes are RR-charged objects. The the exchange by these charges between D-branes is carried out by strings, just as the excitation in an atom is removed by electronic transitions between different levels. So the classification of D-branes acquires a new mathematical interpretation, which will be presented in this paper.

\section{K group calculations}
\par
The problem of D-brane charge classification was raised by Witten, \cite{Witten_1998}. He showed that analyzing of the brane-antibrane system lead to the identification of D-brane charge as an element of the K-theory of the spacetime manifold X as a base for some vector bundles corresponding to D branes. Thus, the interpretation of D-brane charges in terms of K-theory is connected with the basic reason that D-branes carry vector bundles. For IIB string theory one considers a configuration of equal number of D9 and anti-D9-branes carrying vector bundles E and F. The pair (E, F) defines a class in K-theory.
\par
From \cite{Minasian_1997} is known, that wrapped D-branes around supersymmetric cycles $f: W ֒\rightarrow S$ with vector-bundle $E \rightarrow W$, called the Chan-Paton bundle are charged under the RR gauge fields. RR charge of D brane is determined by formula

\begin{equation}
Q = {\rm ch} \left( f_!E \right)\sqrt{\widehat{A} \left( \Tau S \right)},
\end{equation}
where $\Tau S$ is the tangent bundle to spacetime and $f_!$ is the K-theoretic Gysin map. 
\par
But the question is connected with finding of RR charge of D branes in topologically nontrivial B fields. In the presence of the Neveu-Schwarz B-field interacting with D brane the field strength, H, is determined by formula

\begin{equation}
H_{\mu \nu \rho} = \partial B_{\nu \rho} + \partial B_{\rho \mu} + \partial B_{\mu \nu}
\end{equation}

From the paper \cite{Bouwknegt_2000} is well known, that the incorporation of Neveu-Schwarz B-field with three-form field strength $H$ and characteristic class $\left[H\right] \in H^3(X,\mathbb{Z})$ allows to interpret the gauge fields on the D-brane as connections over noncommutative algebras rather than as connections on vector bundles, \cite{Connes_1998}. As the cancelation of global string worldsheet anomalies requires $[H]$ to be a torsion element, the incorporation of nontorsion $[H]$ leads to the limit $n \rightarrow \infty$ of principal $PU(H) = U(H)/U(1)$ bundles over $X$ with $H$ - an infinite dimensional, separable, Hilbert space. For such bundles sections became $C^*$-algebra of continuous sections of the algebra bundle over infinite dimensional, separable, Hilbert space and $C^*$ algebra is itself became Hilbert $A$-module. 

\par
There must be the modifications in consideration of the sections of bundles corresponding to such D branes, \cite{Malyuta_2003}

\begin{equation*}\label{malyuta_eq1}
  \begin{CD}
    SU(n) / \mathbb{Z}_n @>>> P_H  \\
    @. @VVV \\
    @. X
  \end{CD}
\end{equation*}

\begin{equation}\label{malyuta_eq2}
  \begin{CD}
    \lim\limits_{n \to \infty} {SU(n)} / \mathbb{Z}_n @>>> P_H  \\
    @. @VVV \\
    @. X
  \end{CD}
\end{equation}

Isomorphism classes of principal $PU(H)$ bundles over $X$ are parametrized by $H^3(X,\mathbb{Z})$. The upper part of (\ref{malyuta_eq1}) is principal bundle called Azumaya bundle with $n[H] = 0$, where $H_{\mu \nu \lambda}=0$, $B_{\mu \nu} \neq 0$; the lower part of (\ref{malyuta_eq1}) is  principal bundle with $[H] \neq 0$, where $H_{\mu \nu \lambda} \neq 0$, $B_{\mu \nu} \neq 0$ called Rosenberg bundle \cite{Rosenberg_1989}. 
\par
Vector bundles associated with principal one are the following
\begin{equation}
E_H = P_H \times M_c (\mathbb{C}), \; {\rm where} \; Aut\left(M_c (\mathbb{C}) \right) = SU(n) / \mathbb{Z}_n
\end{equation}
\begin{equation}
E_H = P_H \times \mathcal{K}, \; {\rm where} \; Aut\left(\mathcal{K} \right) = \lim\limits_{n \to \infty}SU(n) / \mathbb{Z}_n
\end{equation}
where $M_c (\mathbb{C})$ is $n\times n$ matrix algebra, $\mathcal{K}$ is the algebra of compact operators. It turns out that isomorphism classes of locally trivial bundle $\varepsilon_{[H]}$ over $X$ with fiber $\mathcal{K}$ and structure group $Aut(\mathcal{K} )$ are also parametrized by the cohomology class in $H^3(X,\mathbb{Z})$ called the Dixmier-Douady invariant of $\varepsilon_{[H]}$ and denoted by $\delta \left( \varepsilon_{[H]} \right) = [H]$, $[H] \in H^3(X, \mathbb{Z})$ \cite{BSMF_1963__91__227_0}

\par
As was stressed by Witten in \cite{doi:10.1142/S0217751X01003822}, for two Azumaya bundles, $W$, with string between these twisted bundles, the algebra of $W-W$ open string field theory reduces to the algebra $A_{W(X)}$ of linear transformations of the bundle $W$. In general, $W$ is locally trivial, so $A_{W(X)}$ is isomorphic to $A(X)\otimes M_N$ where $M_N$ is the algebra  of $N\times N$ complex-valued matrices. There is also used the fact that for distinct twisted bundles $W$ and $W’$, the corresponding algebras are "Morita-equivalent" and $K(A_W)=K(A_{W’})$. There $K$-theory is taken in $[H]=0$ case for the noncommutative Azumaya algebra over compact space $X$. As was stressed in \cite{Witten_1998} in the case of Azumaya bundles the groups $K(X)$ and $K(X, [H])$ over compact space $X$  are rationally equivalent.

\par
In most physical applications, \cite{doi:10.1142/S0217751X01003822} for the case of Type IIB string theory  with nontorsion $[H]\neq0$ we have infinite set of $D_9$ or anti $D_9$ branes with infinite rank twisted gauge bundle $E$ or $F$. $D$ brane charge is classified by $K_H$ group of pairs $(E,F)$ modulo the equivalence relation.

\par
So, we can say, that gauge fields on D brane in the presence of B field are interpreted as connections over noncommutative algebras, \cite{Bouwknegt_2000}. Thus, D-brane charges in the presence of B field with nontrivial $[H]$ are classified by K-theory of some noncommutative algebra,  $C^*$-algebra of continuous sections of isomorphic classes of locally trivial bundles $\varepsilon_{[H]}$ over $X$ with fibre $K$ and structure group $PU(H)=Aut(\mathcal{K})$. 

\begin{equation}
K^j(X,[H]) = K_j\left( C_0(X, \varepsilon_{[H]}) \right), \; j=0,1.
\end{equation}
$K$ is the $C^∗$-algebra of compact operators on $H$ - an infinite dimensional, separable, Hilbert space. Therefore, D-brane charges in the presence of a B-field are identified with defined by Rosenberg  twisted K-theory of infinite-dimensional, locally trivial, algebra bundles of compact operators, introduced by Dixmier and Douady. 

\par
The set of all linear operators form a linear space. In particular:
\begin{itemize}
\item the sum of the linear operators and the product of the linear operator by number are determined;
\item the norm of the operator is defined;
\item triangle inequalities are satisfied;
\item the validity of the homogeneity property of the norm is verified.
\end{itemize}

Let $X, Y$ be linear normalized operators. A linear operator $A: X\rightarrow Y$ is said to be bounded if there is a M = const such that

$$
\mid Ax \mid \leq  M \mid x \mid \; \rm{for\: any} \; x \in X.
$$

A classic example of a $C^*$ -algebra is the algebra $B(H)$ of bounded (or equivalent continuous) linear operators defined on a complex Hilbert space $H$.

\par
The classification of algebras with locally compact spectrum $X$ is facilitated by stable isomorphism classes of algebras ( for example, $A$ and $B$ are isomorphic, if $A\otimes K\simeq B\otimes K$) over locally compact Hausdorff space with countable basis of open sets. The reason is connected with the fact that the bundle $\varepsilon_{[H]}$ is the unique locally trivial bundle over $X$ with $\delta \left( \varepsilon_{[H]} \right) = [H]$. There is bijection between isomorphism classes of algebras  whose irreducible representations are infinite-dimensional, "locally trivial" and Cech cohomology group $H^3(X,\mathbb{Z})$. 

\par
To compute $K(A)$ we can use Mayer-Vietoris sequence \cite{Rosenberg_1989}, from which the Dixmier - Douady invariant is determined as  the image in Cech cohomology. 

\begin{equation}
H^2(PU(H),\mathbb{Z}) \rightarrow H^3(X,\mathbb{Z})\rightarrow 0.
\end{equation}
Here $X=Y\cup Z$ , $Y$ and $Z$ are closed subsets of $X$ and $Y\cap Z\rightarrow PU(H)$. The interesting case is connected with $\{Y_n\}$ - some covering of $X$ and $A$ restricts to $C(Y_n)\otimes K$ on $Y_n$ , $n \rightarrow \infty$ and $PU(H)$ is classifying space of line bundles determined for intersections of $Y_n$. Thus for nontorsion case, $[H]\neq 0$, $K_0(A)=0$ and $K_1(A)\cong\mathbb{Z}_n$. 

\par
The same result can be obtained in another way. From \cite{trove.nla.gov.au/work/22701303} it is known, that it can be determined an extension Ext(A,C) of C* algebra A and C by B together with morphisms α and β for which the following sequence is exact
\begin{equation}
\begin{CD}
E: 0 @>>> A @>\alpha >> B @>\beta >> C @>>> 0
\end{CD}
\end{equation}
From long exact sequence of abelian groups of $C^*$ algebras
$$
\begin{CD}
…@>>>Ext_0(A)@>>>Ext(C)@>>>Ext(B)@>>>Ext(A)@>>>…,
\end{CD}
$$

can be received the following result $Ext_0(A)\cong \mathbb{Z}_n$.

\section{Conclusions}
\par
The main purpose af the paper is connected with searches for the RR charge of D branes in the presence of B field. The case of usual D brane RR charge is studied and the answer is known. The presence of the B field is an important physical and mathematical unsolved problem. 
\par
Our task boils down into two issues. First, there is considered $[H]=0$ case for the noncommutative Azumaya algebra over compact space $X$. The algebra of $W-W$ open string field theory between these twisted Azumaya bundles reduces to the algebra $A_{W(X)}$ of linear transformations of the bundle $W$. We also used the fact of "Morita-equivalence" of distinct twisted bundles $W$ and $W'$ and rationally equivalence of the groups $K(X)$ and $K(X, [H])$ over compact space $X$.
\par
The second case is more important, less studied and connected with Rosenberg bundles and with the need to calculate the corresponding K-group. D-brane charges in the presence of B field with nontrivial [H] are classified by K-theory of some noncommutative algebra,  $C^*$-algebra of continuous sections of isomorphic classes of locally trivial bundles. But the description for torsion elements is more natural as the bundle $\varepsilon_{[H]}$ is the unique locally trivial bundle over $X$. We have considered only compact space $X$ and calculated $Ext_0(A)\cong \mathbb{Z}_n$. 
\par
These results can be compared with the corresponding calculations of massless Ramond states of open strings connecting D-branes wrapped on submanifolds of Calabi-Yau's, with holomorphic gauge bundles. The massless Ramond spectra of open strings connecting D-branes are counted by Ext groups and obtained results could be reinterpreted in the language of particles for the corresponding RR charges. The realization of module space in terms of $SU(5)$ multiplets gives supersymmetric matter content \cite{Malyuta_2011}. So, it would be interesting to understand the particle realization for the considered twisted bundles.


\begin{thebibliography}{10}

\bibitem{Bouwknegt_2000}
P.~Bouwknegt and V.~Mathai, ``D-branes, b-fields and twisted k-theory,'' {\em
  Journal of High Energy Physics}, vol.~2000, pp.~007--007, mar 2000.

\bibitem{PhysRevLett.75.4724}
J.~Polchinski, ``Dirichlet branes and ramond-ramond charges,'' {\em Phys. Rev.
  Lett.}, vol.~75, pp.~4724--4727, Dec 1995.

\bibitem{doi:10.1002/zamm.19880680630}
M.~B. Green, J.~H. Schwarz, and E.~Witten, ``Superstring theory. vol. 1:
  Introduction.,'' {\em ZAMM - Journal of Applied Mathematics and Mechanics /
  Zeitschrift für Angewandte Mathematik und Mechanik}, vol.~68, no.~6,
  pp.~258--258, 1988.

\bibitem{Witten_1998}
E.~Witten, ``D-branes and k-theory,'' {\em Journal of High Energy Physics},
  vol.~1998, pp.~019--019, dec 1998.

\bibitem{Minasian_1997}
R.~Minasian and G.~Moore, ``K-theory and ramond-ramond charge,'' {\em Journal
  of High Energy Physics}, vol.~1997, pp.~002--002, nov 1997.

\bibitem{Connes_1998}
A.~Connes, M.~R. Douglas, and A.~Schwarz, ``Noncommutative geometry and matrix
  theory,'' {\em Journal of High Energy Physics}, vol.~1998, pp.~003--003, feb
  1998.

\bibitem{Malyuta_2003}
Y.~Malyuta, ``Nonlinear problems and homological algebra,'' {\em Nonlinear
  boundary problems}, no.~13, pp.~114--117, 2003.

\bibitem{Rosenberg_1989}
J.~Rosenberg, ``Continuous trace algebras from the bundle theoretic point of
  view,'' {\em Jour. Austr. Math. Soc.}, no.~47, pp.~368--381, 1989.

\bibitem{BSMF_1963__91__227_0}
J.~Dixmier and A.~Douady, ``Champs continus d'espaces hilbertiens et de $c^\ast
  $-alg\`ebres,'' {\em Bulletin de la Soci\'et\'e Math\'ematique de France},
  vol.~91, pp.~227--284, 1963.

\bibitem{doi:10.1142/S0217751X01003822}
E.~Witten, ``Overview of k-theory applied to strings,'' {\em International
  Journal of Modern Physics A}, vol.~16, no.~05, pp.~693--706, 2001.

\bibitem{trove.nla.gov.au/work/22701303}
N.~E. Wegge-Olsen, {\em K-theory and C*-algebras : a friendly approach}.
\newblock Oxford ; New York : Oxford University Press, 1993.

\bibitem{Malyuta_2011}
Y.~Malyuta and T.~Obikhod, ``High energy physics and triangulated categories,''
  {\em Ukr. J. Phys.}, vol.~56, no.~5, pp.~411--415, 2011.

\end{thebibliography}
\end{document}